\documentstyle[eqsecnum,pre,preprint,aps,epsfig]{revtex}

\tightenlines

\newcommand{\bq}{\begin{equation}}
\newcommand{\eq}{\end{equation}}
\newcommand{\bqa}{\begin{eqnarray}}
\newcommand{\eqa}{\end{eqnarray}}
\newcommand{\ben}{\begin{enumerate}}
\newcommand{\een}{\end{enumerate}}
\newcommand{\bc}{\begin{center}}
\newcommand{\ec}{\end{center}}
\newcommand{\bqb}{\begin{eqnarray*}}
\newcommand{\eqb}{\end{eqnarray*}}

\begin{document}

\draft
\preprint{PM/04-32,~~December 13, 2004,~~
hep-ph/0410089}

\title{\vspace{1cm}  Electroweak supersymmetric effects on high energy 
unpolarized and polarized single top production at LHC.
\footnote{Partially supported by EU contract HPRN-CT-2000-00149}}
\author{M. Beccaria$^{a,b}$,
F.M. Renard$^c$ and C. Verzegnassi$^{d, e}$ \\
\vspace{0.4cm}
}

\address{
$^a$Dipartimento di Fisica, Universit\`a di
Lecce \\
Via Arnesano, 73100 Lecce, Italy.\\
\vspace{0.2cm}
$^b$INFN, Sezione di Lecce\\
\vspace{0.2cm}
$^c$ Physique
Math\'{e}matique et Th\'{e}orique, UMR 5825\\
Universit\'{e} Montpellier
II,  F-34095 Montpellier Cedex 5.\hspace{2.2cm}\\
\vspace{0.2cm}
$^d$
Dipartimento di Fisica Teorica, Universit\`a di Trieste, \\
Strada Costiera
 14, Miramare (Trieste) \\
\vspace{0.2cm}
$^e$ INFN, Sezione di Trieste\\
}

\maketitle

\begin{abstract}

We consider various processes of  single top production at LHC 
in the theoretical framework of the MSSM and examine the role 
of the supersymmetric electroweak one-loop corrections 
in a special moderately light SUSY
scenario, in an initial parton-pair c.m. 
high energy range  where  a logarithmic asymptotic expansion 
of Sudakov type can be  used. 
We show that the electroweak virtual effects 
are systematically large, definitely beyond the relative ten
percent size, particularly for a final $tH^-$ pair where a 
special enhancement is present. We  show then in a qualitative way
the kind of precision tests of the model that would be 
obtainable from  accurate measurements of the energy distributions 
of the various cross sections and of the top polarization asymmetries.

\end{abstract}

\section{Introduction}

The process of single top production in proton-proton 
collisions has recently been described in detail
\cite{CERNYB} and its relevance in  theoretical 
models of electroweak physics has been stressed
with special emphasis. In the particular framework of the 
Standard Model, the relevant final pairs can be
(at partonic level): 1) a top and a light quark ($t$-channel 
production), 2) a top and a $\bar b$ ($s$-channel production),
 3) a top 
and a $W^-$ (associated production), and the corresponding 
Feynman diagrams at Born level are shown in Fig.1,2.
As stressed in Ref.\cite{CERNYB}, from precise measurements of these 
processes the first direct determination of the CKM $Wtb$
coupling at hadron colliders will be obtained, which already 
motivates the existence of dedicated experimental and theoretical
studies.\par
The theoretical analysis of the first two processes 
has been performed in a number of papers, both within the
SM  and  beyond it, with special emphasis on the 
production at TEVATRON \cite{TEVAT}. 
The third process
will only be measured at LHC, and to our knowledge 
a detailed theoretical study at the perturbative one-loop level,
valid in particular for the MSSM, does not exist yet.\par
In this paper, we shall assume a preliminary discovery 
of supersymmetry at LHC (perhaps already at TEVATRON)
and a SUSY scenario of a "moderately" light kind, 
in which all sparticle masses lie below, say, approximately three-four 
hundred GeV. Our aim will be that of showing that, 
under these conditions, single top production at LHC might
provide crucial, accurate precision tests of the candidate 
theoretical model. For this preliminary analysis we shall stick
to the  MSSM, but our treatment could be modified 
in a straightforward way to examine less simple
supersymmetric proposals.\par
In our analysis, we shall consider "large" 
values of the initial parton pair c.m. energy $\sqrt{s}$, 
typically of the
1 TeV size. There exist two main reasons that motivate 
our choice. The first one is that within this energy range
we shall feel entitled to make use, at the one-loop 
perturbative level, of  a simple asymptotic logarithmic 
expansion of
so called Sudakov kind, whose validity in the MSSM 
is related to the assumption of a light SUSY scenario, so that
$s/M^2$ (the typical parameter of the logarithmic expansion) 
is of order ten ($M$ is the heaviest SUSY mass of the
scenario). The second one is that, for $\sqrt{s}$ values much 
larger than the final $t$, $W$ masses, remarkable 
kinematical simplifications arise whose 
theoretical origin will be shown. These make, in particular,
the treatment of the associated production process much 
simpler, and allow to understand the potentialities of its
relevant supersymmetric effect  from inspection of  
short analytic formulae.\par
The plan of this paper will be the following. 
In Section 2A we shall briefly derive the expressions 
of the electroweak
Sudakov expansions at one loop to next-to leading order, 
i.e. retaining the squared and the linear
 logarithmic terms, for the $t$- and $s$-channel processes. 
For the latter, a partial MSSM calculation of the Yukawa effect 
is already available,
in an arbitrary c.m. energy configuration where in 
general all the parameters of the model will enter the theoretical
predictions~\cite{Yukawa}; our calculation will include logarithmic terms
of several origins, but owing to our energy choice a 
strongly reduced number of 
parameters will remain in the relevant expressions, as we shall show
in detail. Section 2B will be devoted to a more 
accurate description of the determination of the 
Sudakov expansion for
the so called associated $tW$ production process. 
Here, as we said,
a theoretical calculation in the MSSM is not known 
to us; independently of this, we shall comment anyway a few
special features of our expansions, in particular the simplifications due 
to the choice of a high energy regime.
Within the MSSM, another process to 
be added to the previously considered ones seems 
to us to be that of
$tH^-$ production. This has also been already 
studied in an arbitrary energy configuration 
\cite{BGGS},\cite{Berger},\cite{RM}; we shall derive here in 
Section 2C only its electroweak logarithmic 
Sudakov expansion, to be compared with the three 
previous ones to
have a more complete description of the asymptotic 
single top production at LHC.

A final topic that seems to us to deserve some consideration
is that of the single top polarization. This is automatically fixed in the 
$t$, $s$ channel processes in the chosen high energy region where the 
final top must be of left-handed type; this feature is not valid for $tW^-$
and $tH^-$ production. As a consequence, one may in principle consider 
longitudinal polarization asymmetries for these processes.
In Section 2D, we shall derive the asymptotic expressions of these 
quantities, that will turn out to be relatively simple.

Having at disposal all the relevant asymptotic 
expressions for the four considered processes, we shall then
devote Section 3 to an investigation of which 
kind of precision test of the model could be provided by
realistic experimental measurements. This will 
be done in a necessarily qualitative way, exploiting the
available preliminary conclusions of Ref.\cite{CERNYB}, 
assuming that the distributions in $\sqrt{s}$ can be determined with a
certain accuracy in our chosen range. 
Our presentation will remain for the moment
at a qualitative level. In fact, the purpose of our 
paper is mostly that of
showing the possible relevance of measurements 
that appear to us, at least in principle, performable, and
to encourage an experimental performance like 
that indicated in Ref.\cite{CERNYB}. This point will be 
discussed in the final
conclusions, given in the short Section 4.

\section{Single top production}

As we said, this Section will be devoted to a derivation 
of the asymptotic electroweak expressions for the
four considered processes. To save space and time, 
since the major part of the relevant technical definitions and
properties has been already exhaustively illustrated 
in previous papers~\cite{Sudakov}, 
we shall quickly write
the final formulae for the ($t,d$), ($t,\bar b$) and ($t,H^-$) processes, 
that have already been
studied in the literature (but not in our chosen scenario) 
as we anticipated. The treatment of the final ($t,W^-$) pair will be
slightly more detailed since for this process, as we said, we are not aware 
of a theoretical  investigation in the
MSSM, and also since in this case the benefits of the chosen 
high energy configuration deserve, in our opinion, a short
mention. 

\subsection{T,S CHANNELS  SINGLE TOP PRODUCTION}

\subsubsection{$bu\to td$ with $W$ exchange in the $t$ channel}

We begin with the $t$-channel process represented in 
Fig.1a  at Born level. As one sees, there  is only
one diagram with W  exchange in the $t$-channel, $t=(p_b-p_t)^2$. 
The corresponding scattering amplitude is:

\bq
A^{Born}={2\pi\alpha\over s^2_W(t-m^2_W)}
[\bar u(d,\tau')\gamma^{\mu}P_Lu(u,\lambda')]
[\bar u(t,\tau)\gamma_{\mu}P_Lu(b,\lambda)]
\eq
\noindent
where $\lambda,\lambda',\tau,\tau'$ are the $b,u,t,d$
helicities,
$P_{R,L}=(1\pm\gamma^5)/2$ are the projectors on
$R,L$ chiralities ans $s_W$ is the sine of the Weinberg angle.

It  is convenient to work with 
helicity amplitudes $F_{\lambda,\lambda',\tau,\tau'}$; 
retaining only the top 
mass and setting
all the remaining masses equal to zero leaves one single 
amplitude $F_{----}$:

\bq
F^{Born}_{----}={4\pi\alpha s \sqrt{\beta}\over s^2_W(t-m^2_W)}
\eq
\noindent
with $\beta={p_t\over E_t}=1-{m^2_t\over s}$.

The expression of the differential cross section is:

\bq
{d\sigma\over dcos\theta}={\beta
\over1152\pi s}\sum_{spin,col}|F_{\lambda \lambda'\tau\tau'}|^2
\eq

which becomes after color average:

\bq
{d\sigma^{Born}\over d\cos\theta}=
{\beta^2\pi\alpha^2 s\over8s^4_W(t-m^2_W)^2}
\eq

At one-loop, the Sudakov electroweak corrections can 
be of universal and of angular dependent kind, 
and we follow the definitions of ~\cite{Sudakov}.
 
The effect of the universal terms on the helicity amplitude
can be summarized as follows:

\bq
F^{Univ}_{----}=
F^{Born}_{----}
~{1\over2}~[~c^{ew}(b\bar b)_L+c^{ew}(u\bar u)_L+
c^{ew}(d\bar d)_L+c^{ew}(t\bar t)_L]
\eq

where:

\bq
c^{ew}(q\bar q)_L=c^{ew}(\tilde{q}\tilde{\bar q})_L=
c(q\bar q, ~gauge)_L~+~c(q\bar q,~yuk)_L
\eq

\bq
c(d\bar d, ~gauge)_L=c(u\bar u, ~gauge)_L={\alpha(1+26c^2_W)\over144\pi
s^2_Wc^2_W}~(2~ \log{s\over m^2_W}-\log^2{s\over m^2_W})
\eq
\bq
c(b\bar b, ~yuk)_L=
c(t\bar t, ~yuk)_L=-~{\alpha\over8\pi s^2_W}~
[\log{s\over m^2_W}]~
[{m^2_t\over m^2_W}(1+\cot^2\beta)+{m^2_b\over m^2_W}(1+\tan^2\beta)]
\eq

Note that we use $m_W$ as a scale for the log term. This is mandatory
for the squared log which arises from $W$ loops. For what concerns the
single log, as we do not consider "constant terms" this a matter of
choice and we decide to keep $m_W$ as a scale for the above terms.\par
The   angular dependent terms have the following expression:

\bq
F^{ang}_{----}=
F^{Born}_{----}
[~c^{ang}_{----}~]
\eq

\bqa
c^{ang}_{----}&=&-~{\alpha(1+8c^2_W)\over18\pi s^2_Wc^2_W} 
[\log{-u\over s}][\log{s\over m^2_W}]
\nonumber\\
&&
-~{\alpha(1-10c^2_W)\over36\pi s^2_Wc^2_W} 
[\log{-t\over s}][\log{s\over m^2_W}]
\eqa
At high energy we have $t\simeq-~{s\over2}(1-\cos\theta)$
and $u\simeq-~{s\over2}(1+\cos\theta)$.\par

There are also SUSY QCD corrections, only of universal 
kind. We shall treat them as "known" terms  for
what concerns our search of electroweak effects;
at one loop, their expression is not difficult to be derived, 
and reads:

\bqa
F^{Univ~SUSYQCD}_{----}&=&
F^{Born}_{----}
~[-\frac{2\alpha_s}{3\pi}\log\frac{s}{M_{\rm SUSY}^2}]
\eqa

In addition to the previous terms of Sudakov type, 
there are at one-loop "known" linear 
logarithms of RG origin, whose expression we quote for completeness:

\bqa
F^{RG}_{----}&=&-~{1\over4\pi^2}[g^4\tilde{\beta^0}
{dF^{Born}_{----}\over dg^2}][\log\frac{s}{M_W^2}]\nonumber\\
&&={\alpha^2 s \sqrt{\beta}\over s^4_W(t-m^2_W)}[\log\frac{s}{M_W^2}]
\eqa

using the lowest order Renormalization
Group $\beta$ function for the gauge
coupling $g=e/s_W$: $\tilde{\beta^0}=-~{1\over4}$ in MSSM.

\subsubsection{$u\bar d\to t\bar b$ with $W$ exchange in the $s$ channel}

Neglecting  again all quark masses with the exception 
of $m_t$ one finds from the Born diagram of Fig. 1b:

\bq
A^{Born}={e^2\over2s^2_W(s-m^2_W)}
[\bar v(\bar d,\lambda')\gamma^{\mu}P_Lu(u,\lambda)]
[\bar u(t,\tau)\gamma_{\mu}P_Lv(\bar b,\tau')]
\eq

There are now in principle two helicity amplitudes $F_{\lambda, \lambda', \tau, \tau'}$:

\bq
F^{Born}_{-+-+}=-~{2\pi\alpha s\sqrt{\beta}(1+\cos\theta)
\over s^2_W(s-m^2_W)}
\eq
\bq
F^{Born}_{-+++}=-~{2\pi\alpha m_t\sqrt{s}\sqrt{\beta}\sin\theta
\over s^2_W(s-m^2_W)}
\eq
\noindent
where $\beta=1-{m^2_t\over s}$.

Note that the second amplitude $F^{Born}_{-+++}$ is depressed  
with respect to the first one  by the factor $m_t/\sqrt{s}$, so that
in our chosen configuration it will be treated with some 
approximations, neglecting its one loop corrections.
 
In the differential cross section
the two contributions will be:

\bq
{d\sigma\over d\cos\theta}={\beta\over128\pi s}
\sum_{spins}|F_{\lambda,\lambda',\tau,\tau'}|^2
\eq

\bq
{d\sigma^{Born}\over d\cos\theta}=
{\beta^2\pi\alpha^2
s\over32s^4_W(s-m^2_W)^2}[(1+\cos\theta)^2+{m^2_t\over s}\sin^2\theta]
\eq

where the first term comes from $F_{-+-+}$ and the second one comes from 
the ``depressed'' amplitude $F_{-+++}$ that will be retained in Born approximation.
For the relevant amplitude $F_{-+-+}$, we obtain the
following effects:

\bqa
F^{Univ}_{-+-+}&=&
F^{Born}_{-+-+}
~{1\over2}~[~c^{ew}(b\bar b)_L+c^{ew}(u\bar u)_L+
c^{ew}(d\bar d)_L+c^{ew}(t\bar t)_L]
\eqa

 with the expressions of the various $c_{ew}$ given for 
the $t$-channel case and

\bqa
F^{ang}_{-+-+}&=&
{\alpha^2(1+cos\theta)\over4s^2_W}[\log\frac{s}{M_W^2}]
\{[4(Q_dQ_b+Q_{u}Q_{t})
+{g^Z_{dL}g^Z_{bL}+g^Z_{uL}g^Z_{tL}\over s^2_Wc^2_W}]~\log{-t\over s}]
\nonumber\\
&&-[4(Q_dQ_{t}+Q_{u}Q_{b})
+{g^Z_{dL}g^Z_{tL}+g^Z_{uL}g^Z_{bL}\over s^2_Wc^2_W}]~\log{-u\over s}]
\}
\eqa

with $g^Z_{qL}=2I^3_q-2s^2_WQ_q$.\\

We list also the SUSY QCD logarithms:

\bq
F^{Univ~SUSYQCD}_{-+-+}=
F^{Born}_{-+-+}
~[-\frac{2\alpha_s}{3\pi}\log\frac{s}{M_{\rm SUSY}^2}]
\eq

and those of RG origin:

\bqa
F^{RG}_{-+-+}&=&-~{1\over4\pi^2}[g^4\tilde{\beta^0}
{dF^{Born}_{-+-+}\over dg^2}]\ \log\frac{s}{M_W^2}\nonumber\\
&&=-~{s\ \alpha^2 \sqrt{\beta}(1+\cos\theta) \over 2s^4_W(s-m^2_W)}\ \log\frac{s}{M_W^2}
\eqa

using $\tilde{\beta^0}=-~{1\over4}$ in  MSSM and choosing $M_W$ as the RG scale,  which seems to
us justified  at our logarithmic level.

For the $t$- and $s$-channel processes, the logarithmic 
effects at one-loop are  fully  summarized by 
eqs.(2.5-2.12, 2.18-2.21).
They represent an original result of this paper, 
and provide an effective simple representation in
the scenario that we have chosen ,valid, we repeat, 
to next-to leading logarithmic order. As one sees, the
welcome feature of the formulae is that in the 
logarithmic expansion there appears only 
one SUSY parameter, $\tan\beta$. 
This would allow a remarkably simplified 
test of the model, that will be discussed in the
next Section 3.

\subsection{ASSOCIATED \lowercase{t},W PRODUCTION: $b\ g\to t\ W^-$}

We now move to the process to which we devote, for various reasons, a more
detailed description in our scheme,
i.e. the so
called associated ($t,W$) production. 
Fig.2 shows that now there are two Born diagrams, 
one ($a$) in the $s$-channel with
exchange of a bottom quark and one ($b$) 
in the $u=(p_b-p_W)^2$ channel with exchange of a top quark. 
Denoting by
$p=|\overrightarrow{p_t}|=|\overrightarrow{p_W}|$ 
the modulus  of the final c.m. momentum, 
we define:

\bq
\beta={2p\over\sqrt{s}}~~~~\beta^{'0}={2p^0_W\over\sqrt{s}}
~~~r_t={p\over E_t+m_t}
=\sqrt{{(\sqrt{s}-m_t)^2-m^2_W\over(\sqrt{s}+m_t)^2-m^2_W}}\eq

and the W coupling $-g_{WL}(\gamma_{\rho}P_L\epsilon^{\rho}_V)$
with $g_{WL}={e\over s_W\sqrt{2}}$.\\

  Neglecting all the quark masses with the 
exception of the top one, one gets the helicity amplitudes 
$F_{\lambda,\mu,\lambda',\mu'}$
with the helicities $\lambda(b)=\pm 1/2$, $\mu(g)=\pm 1$, $\lambda'(t)=\pm 1/2$,
$\mu'(W)\equiv\mu' = \pm 1$ for transverse $W_T^-$ or $\mu'(W) = 0$ for longitudinal 
$W_0^-$ respectively.

\bqa
F^{Born~a}_{\lambda,\mu,\lambda',\mu'}&=&{1\over2}
g_{WL}g_s({\lambda^l\over2})
\sqrt{{(\sqrt{s}+m_t)^2-m^2_W\over s}}\delta_{\lambda,L}~\{\nonumber\\
&&{1+r_t\over2}\cos{\theta\over2}
[1+\mu\mu'\cos\theta+2\lambda(\mu
+\mu'\cos\theta)+2\mu'(\mu+2\lambda)\sin^2{\theta\over2}]
\delta_{\lambda,\lambda'}
\nonumber\\
&&-~{1-r_t\over2}\sin{\theta\over2}
[2\lambda(1+\mu\mu'\cos\theta)+\mu
+\mu'\cos\theta-2\mu'(1+2\lambda\mu)\cos^2{\theta\over2}]
\delta_{\lambda,-\lambda'}~\}
\nonumber\\
&&\eqa

\bqa
F^{Born~a}_{\lambda,\mu,\lambda',0}&=&
g_{V\lambda}g_s({\lambda^l\over2})
({\sqrt{(\sqrt{s}+m_t)^2-m^2_W}\over2M_W\sqrt{2}})
\delta_{\lambda,L}\nonumber\\
&&
\{~{1+r_t\over2}(\mu+2\lambda)\sin{\theta\over2}
(\beta+\beta^{'0})\delta_{\lambda,\lambda'}\nonumber\\
&&+{1-r_t\over2}(1+2\lambda\mu)\cos{\theta\over2}
(\beta-\beta^{'0})\delta_{\lambda,-\lambda'}~\}
\eqa

\bqa
F^{Born~b}_{\lambda,\mu,\lambda',\mu'}&=&
{g_{WL}g_s s\over4(u-m^2_t)}({\lambda^l\over2})
\sqrt{{(\sqrt{s}+m_t)^2-m^2_W\over s}}\delta_{\lambda,L}~\{~
\nonumber\\
&&{1+r_t\over2}
\{(1-\beta^{'0})(1+\mu\mu'(2\cos\theta-1)
-2\lambda(\mu+\mu'))\nonumber\\
&&-(1-\cos\theta)(\mu\mu'+\beta)
-\mu\mu'(\beta+\cos\theta)-(1+\beta \cos\theta)\nonumber\\
&&
+2\lambda(\mu(1+\beta cos\theta)+\mu'(\beta+ \cos\theta)
-(1-\cos\theta)(\mu'+\mu\beta))\}\cos{\theta\over2}
\delta_{\lambda,\lambda'}
\nonumber\\
&&+{1-r_t\over2}
\{(1-\beta^{'0})(-2\mu'\cos^2{\theta\over2}+\mu+\mu'\cos\theta-
2\lambda(2\mu\mu'\cos^2{\theta\over2}+1+\mu\mu'\cos\theta))\nonumber\\
&&
-\mu(1+\beta \cos\theta)-\mu'(\beta+\cos\theta)-2\cos^2{\theta\over2}
(\mu\beta+\mu')-4\lambda(\mu\mu'+\beta)\cos^2{\theta\over2}\nonumber\\
&&
+2\lambda\mu\mu'(\beta+\cos\theta)+2\lambda(1+\beta \cos\theta)\}
\sin{\theta\over2}
\delta_{\lambda,-\lambda'}\nonumber\\
&&+{2m_t\over\sqrt{s}}[\delta_{\lambda,\lambda'}{1-r_t\over2}
\cos{\theta\over2}(-1+\mu\mu'(1-2\cos\theta)+2\lambda(\mu+\mu'))
\nonumber\\
&&+\delta_{\lambda,-\lambda'}{1+r_t\over2}\sin{\theta\over2}
(\mu'-\mu+2\lambda(1+\mu\mu'+2\mu\mu'\cos\theta))]
~\}
\eqa

\bqa
F^{Born~b}_{\lambda,\mu,\lambda',0}&=&
g_{WL}g_s({\lambda^l\over2})
({s\sqrt{(\sqrt{s}+m_t)^2-m^2_W}\over4M_W(u-m^2_t)\sqrt{2}})
\delta_{\lambda,L}\nonumber\\
&&
\{~{1+r_t\over2}\sin{\theta\over2}(1+\cos\theta)2\beta
[\mu(\beta-\beta^{'0}+1)+2\lambda]\delta_{\lambda,\lambda'}\nonumber\\
&&
+{1-r_t\over2}\cos{\theta\over2}2\beta[(1+2\lambda\mu \cos\theta)
(1+\beta-\beta^{'0})+2\lambda\mu+\cos\theta]
\delta_{\lambda,-\lambda'}
\nonumber\\
&&+{2m_t\over\sqrt{s}}[\delta_{\lambda,\lambda'}{1-r_t\over2}
\sin{\theta\over2}(\beta^{'0}(2\lambda-\mu(1+2\cos\theta))
-\beta(2\lambda+\mu))
\nonumber\\
&&+\delta_{\lambda,-\lambda'}{1+r_t\over2}\cos{\theta\over2}
(-\beta^{'0}(1-2\lambda\mu(1-2\cos\theta)-\beta(1+2\lambda\mu))]~\}
\eqa
where $\lambda^l$ are the color matrices associated with the gluon and $g_s$
is the QCD coupling constant.

With the color sum 
$\sum_lTr[({\lambda^l\over2})({\lambda^l\over2})]=4$
the cross section is

\bq
{d\sigma\over d\cos\theta}={\beta\over768\pi s}
\sum_{spins}|F_{\lambda,\mu,\lambda',\mu'}|^2
\eq

In our special scenario we are allowed to neglect $m_t/\sqrt{s}$,
$m_W/\sqrt{s}$, but not $m_t/M_W$. This leads to a remarkable
simplification of the previous expressions, that become now:

\bqa
F^{Born~a}_{\lambda,\mu,\lambda,\mu}&\to&
g_{WL}g_s({\lambda^l\over2})
\cos{\theta\over2}(1+2\lambda\mu)\delta_{\lambda,L}
\eqa

\bqa
F^{Born~b}_{\lambda,\mu,\lambda,\mu}&\to&g_{WL}g_s{s\over u}
({\lambda^l\over2})
\cos{\theta\over2}(2\lambda\mu \cos\theta-1)\delta_{\lambda,L}
\eqa

\bqa
F^{Born~a}_{\lambda,\mu,\lambda,0}&\to&
g_{WL}g_s({\lambda^l\over2})\delta_{\lambda,L}
{\sqrt{s}\over\sqrt{2}M_W}\sin{\theta\over2}(\mu+2\lambda)
\eqa

\bqa
F^{Born~b}_{\lambda,\mu,\lambda',0}&\to&
-g_{WL}g_s({\lambda^l\over2})\delta_{\lambda,L}
\{~{\sqrt{s}\over\sqrt{2}M_W}\sin{\theta\over2}(\mu+2\lambda)
\delta_{\lambda,\lambda'}+{m_t\over\sqrt{2}M_W}\cos{\theta\over2}
(1+2\lambda\mu)\delta_{\lambda,-\lambda'}\nonumber\\
&&
-{\sqrt{2}m_t\over M_W}\delta_{\lambda,-\lambda'}
\cos{\theta\over2}[{1+2\lambda\mu \cos\theta\over
1+\cos\theta}]~\}
\eqa

Explicitly, the only remaining amplitudes are

\bq
F^{Born~a+b}_{----}\to g_{WL}g_s({\lambda^l\over2})
{2\over \cos{\theta\over2}}
\eq

\bq
F^{Born~a+b}_{-+-+}\to g_{WL}g_s({\lambda^l\over2})
{2 \cos{\theta\over2}}
\eq

\bq
F^{Born~a+b}_{-++0}\to g_{WL}g_s({\lambda^l\over2})
{\sqrt{2}m_t\over M_W}
\cos{\theta\over2}({1-\cos\theta\over1+\cos\theta})
\label{fbornl}\eq

They provide a rather simple expression 
of the differential cross section:

\bq
{d\sigma^{Born}\over d\cos\theta}\to
-~{\pi \alpha\alpha_s\over24s^2_Wus^2}[s^2+u^2+{m^2_tt^2\over2M^2_W}]
\eq

The electroweak  Sudakov terms at the one-loop level 
are now relatively simple to compute and to show. More
precisely, we obtain for the universal component for transverse $W^-_T$ production:

\bqa
F^{Univ}_{-,\mu,-,\mu}&=&
F^{Born}_{-,\mu,-,\mu}
[~{1\over2}~(~c^{ew}(b\bar b)_{L}
+c^{ew}(t\bar t)_{L}~)~+c^{ew}(W_T)]
\eqa
\noindent
where $c(b\bar b)$, $c(t\bar t)$ have been already 
defined by eqs.(2.6-2.8) (following the notations
of~\cite{Sudakov,WWpaper}) and 
\bq
c^{ew}(W_T)={\alpha\over4\pi s^2_W}[-\log^2\frac{s}{M_W^2}]
\eq

and for  longitudinal $W^-_0$ production:

\bqa
F^{Univ}_{-,+,+,0}&=&
F^{Born}_{-,+,+,0}
[~{1\over2}~(~c^{ew}(b\bar b)_{L}
+c^{ew}(t\bar t)_{R}~)~+c^{ew}(W_0)]
\label{funivl}
\eqa
where in MSSM

\bqa
c^{ew}(W_0)&=&{\alpha\over4\pi}\{
[-~{1+2c^2_W\over8s^2_Wc^2_W}\log^2\frac{s}{M_W^2}]\nonumber\\
&&+[\log\frac{s}{M_W^2}][-~{17+10 c^2_W\over36s^2_Wc^2_W}+
{m^2_b\over4s^2_WM^2_W}(1+\tan^2\beta)+
{3m^2_t\over4s^2_WM^2_W}(1+\cot^2\beta)]\}\nonumber\\
&&
\eqa

such that
\bqa
F^{Univ}_{-,+,+,0}&=&
F^{Born}_{-,+,+,0}[~{\alpha\over4\pi}]\{~[-\log^2\frac{s}{M_W^2}]
[{13+14c^2_W\over36s^2_Wc^2_W}]~\}
\label{funivlsm}
\eqa

For the  electroweak angular terms we find:

\bqa
F^{ang}_{-,\mu,-,\mu}&=&
F^{Born}_{-,\mu,-,\mu}
[-~{\alpha\over2\pi}][\log\frac{s}{M_W^2}]\{~[\log{-t\over s}]
[Q_bQ_{t}+{g^Z_{bL}g^Z_{tL}\over4s^2_Wc^2_W}]
+~{1\over s^2_W}\log{-u\over s}~\}\nonumber\\
&&
=F^{Born}_{-,\mu,-,\mu}[-~{\alpha\over2\pi}][\log\frac{s}{M_W^2}]\{~[\log{-t\over s}]
[{1-10c^2_W\over36s^2_Wc^2_W}]
+~{1\over s^2_W}\log{-u\over s}~\}\nonumber\\
&&
\label{fangt}\eqa

\bqa
F^{ang}_{-,+,+,0}&=&
F^{Born}_{-,+,+,0}[-~{\alpha\over24\pi c^2_W}][\log\frac{s}{M_W^2}]
\{~[\frac{4}{3} \log{-t\over s}]-{1-10c^2_W\over s^2_W}\log{-u\over s}~\}
\label{fangl}\eqa

There are also SUSY QCD universal terms:

\bqa
F^{Univ~SUSYQCD}_{-,\mu,-,\mu}&=&
F^{Born}_{-,\mu,-,\mu}
[-\frac{\alpha_s}{3\pi}\log\frac{s}{M^2_{\rm SUSY}}]
\eqa

\bqa
F^{Univ~SUSYQCD}_{-,+,+,0}&=&
F^{Born}_{-,+,+,0}
[-\frac{\alpha_s}{3\pi}\log\frac{s}{M^2_{\rm SUSY}}]
\eqa
\noindent
For this process, there are no one loop ew RG terms.\par
A few comments are now appropriate. 
The first one is that we have obtained a great 
lot of simplification neglecting
systematically terms of order $m_t/\sqrt{s}$ 
in the components of the helicity amplitudes. 
This is in agreement with the general
strategy of our asymptotic scenario, 
that assumes $\sqrt{s}$ to be sufficiently larger 
than all the masses of the
process and will retain only the two leading 
logarithmic terms of the Sudakov expansion. For a different
scenario, e.g. one where $\sqrt{s}$ is smaller, 
the remaining components of the cross section 
must be retained and
the theoretical formulae will be more complicated. 
Their computation might be necessary if for instance
the (optimistic?) assumption of light 
SUSY were not verified. Then the region of large $\sqrt{s}$ would have
no special reasons (of Sudakov origin) 
to be priviledged, and thus a complete 
one-loop calculation would be
imperative, to be used for instance in 
a relative low $\sqrt{s}$ range where a 
Sudakov expansion would certainly not be
acceptable. 
This can certainly be done, but  in our optimistic attitude we
will now show which interesting features might 
arise from a measurement of the process at LHC, if our chosen
scenario were ( kindly) chosen by Nature. 
The second point concerns Eqs.\ref{fbornl},\ref{funivl},2.39,\ref{fangl}. 
As one can verify, they are in full agreement with the equivalence
theorem which states that, neglecting $M_W^2/s$ terms, the expressions for 
longitudinal $W^-_0$ production are identical to those for the Goldstone boson $G^-$.
This can be seen by taking the expressions written in the next Section 2C for $tH^-$ production and 
replacing $M^2_H$ with $M^2_G = M^2_W$, $m_t\cot\beta$ with $m_t$, and 
$m_b\tan\beta$ with $-m_b$, and this constitutes a positive check of our calculation.
 
In fact, having concluded the study  of  
the production in the MSSM of a top and a W boson, 
it seems almost natural to
enlarge the treatment so as to include 
a process which is tightly connected with it, 
i.e. that of production of a top and of
the charged Higgs of the model. 
This process has been already studied~\cite{BGGS,Berger,RM}. 
In particular, in~\cite{Berger} one can find
an exhaustive discussion of the SUSY QCD corrections, particularly relevant in the large
$\tan\beta$ limit. 
In the next Subsection 2C we shall show that the purely electroweak supersymmetric
corrections considered in \cite{BGGS}, 
but not treated in~\cite{Berger},
are also very important especially in the electroweak Sudakov regime
of the LHC in which we are interested.

\subsection{ASSOCIATED \lowercase{t}, $H^-$ PRODUCTION: $bg\to tH^-$}

This process is similar to the longitudinal part of the previous
process $bg\to tW^-$.

There are two Born diagrams like in Fig.2, 
but replacing the final $W^-$
by the $H^-$, (a) $s$-channel with bottom quark exchange
and (b) $u$-channel with top quark exchange. We use the same notations
as before with
$u=(p_b-p_H)^2$ and define:

\bq
\beta={2p_t\over\sqrt{s}}~~~~\beta^{'0}
={2p^0_{H}\over\sqrt{s}}
~~~r_t={p_t\over E_t+m_t}
=\sqrt{{(\sqrt{s}-m_t)^2-m^2_{H}\over(\sqrt{s}+m_t)^2-m^2_{H}}}
\eq
\noindent
and the $H^-tb$ coupling
${e\over\sqrt{2}s_WM_W}[m_t \cot\beta P_L+m_b \tan\beta P_R]$.

Neglecting the quark masses, except $m_t$,
one gets the helicity amplitudes $F_{\lambda,\mu,\lambda'}$
with the $\lambda(b),\mu(g),\lambda'(t)$ 
helicities respectively:

\bqa
F^{Born~a}_{\lambda,\mu,\lambda'}&\to&
{eg_s\over s_WM_W}({\lambda^l\over2})
\sqrt{{(\sqrt{s}+m_t)^2-m^2_{H}\over s}}
\{~2\lambda({1-r_t\over2})\sin{\theta\over2}
\delta_{\lambda,\lambda'}\nonumber\\
&&
+({1+r_t\over2})cos{\theta\over2}
\delta_{\lambda,-\lambda'}~\}\delta_{\lambda,\mu}
[m_t \cot\beta \delta_{\lambda,L}+m_b \tan\beta \delta_{\lambda,R}]
\eqa

\bqa
F^{Born~b}_{\lambda,\mu,\lambda'}&\to&
{eg_s\sqrt{s}\over s_WM_W(u-m^2_t)}({\lambda^l\over2})
\sqrt{{(\sqrt{s}+m_t)^2-m^2_{H}\over s}}
\{~m_t\delta_{\lambda,\mu}[({1+r_t\over2})\sin{\theta\over2}
2\lambda\delta_{\lambda,\lambda'}\nonumber\\
&&+({1-r_t\over2})
\cos{\theta\over2}\delta_{\lambda,-\lambda'}]\nonumber\\
&&+({1-r_t\over2})\sin{\theta\over2}
\delta_{\lambda,\lambda'}[-(2\lambda)p_t(1+\cos\theta)
\delta_{\lambda,-\mu}+(2\lambda)(p_t\cos\theta-E_t+\sqrt{s})
\delta_{\lambda,\mu}]\nonumber\\
&&+({1+r_t\over2})\cos{\theta\over2}
\delta_{\lambda,-\lambda'}[p_t(1-\cos\theta)
\delta_{\lambda,-\mu}+(p_t\cos\theta-E_t+\sqrt{s})
\delta_{\lambda,\mu}]~\}\nonumber\\
&&
[m_t \cot\beta \delta_{\lambda,L}+m_b \tan\beta \delta_{\lambda,R}]
\eqa

With the color sum $\sum_l({\lambda^l\over2})({\lambda^l\over2})=4$
the cross section is

\bq
{d\sigma\over d\cos\theta}={\beta\over768\pi s}
\sum_{spins}|F_{\lambda,\mu,\lambda'}|^2
\eq

neglecting $m_t/\sqrt{s},m_H/\sqrt{s}$, but not 
$m_t/m_W$ nor $m_b/M_W$ we have

\bqa
F^{Born~a+b}_{\lambda,\mu,\lambda'}&\to&
-~{eg_s\sqrt{s}\over s_WM_W}({\lambda^l\over2})\cos{\theta\over2}
({1-\cos\theta\over1+\cos\theta})\delta_{\lambda,-\lambda'}
\delta_{\lambda,-\mu}
[m_t \cot\beta \delta_{\lambda,L}+m_b \tan\beta \delta_{\lambda,R}]
\nonumber\\
&&\eqa

explicitly

\bqa
F^{Born~a+b}_{-,+,+}&\to&
-~{eg_sm_t \cot\beta\over s_WM_W}
({\lambda^l\over2})\cos{\theta\over2}
({1-\cos\theta\over1+\cos\theta})
\eqa
\bqa
F^{Born~a+b}_{+,-,-}&\to&
-~{eg_sm_b \tan\beta\over s_WM_W}
({\lambda^l\over2})\cos{\theta\over2}
({1-\cos\theta\over1+\cos\theta})
\eqa

\bq
{d\sigma^{Born}\over d\cos\theta}\to
-~{\pi\alpha\alpha_s(m^2_t \cot^2\beta+m^2_b \tan^2\beta)
 t^2\over48s^2_WM^2_Wus^2}
\eq

Moving to one loop, one finds in our scenario:

\underline{a) One loop electroweak universal terms}

\bqa
F^{Univ}_{-,+,+}&=&
F^{Born}_{-,+,+}
[~{1\over2}~(~c^{ew}(b\bar b)_{L}
+c^{ew}(t\bar t)_{R}~)~+c^{ew}_{-,+,+}(H^-)]
\eqa

\bqa
F^{Univ}_{+,-,-}&=&
F^{Born}_{+,-,-}
[~{1\over2}~(~c^{ew}(b\bar b)_{R}
+c^{ew}(t\bar t)_{L}~)~+c^{ew}_{+,-,-}(H^-)]
\eqa

with 
\bqa
c^{ew}_{-,+,+}(H^-)&=&{\alpha\over4\pi}\{
[-~{1+2c^2_W\over8s^2_Wc^2_W}\log^2\frac{s}{M_W^2}]\nonumber\\
&&+[\log\frac{s}{M_W^2}][-~{17+10 c^2_W\over36s^2_Wc^2_W}+
{m^2_b\over4s^2_WM^2_W}(1+\tan^2\beta)+
{3m^2_t\over4s^2_WM^2_W}(1+\cot^2\beta)]\}\nonumber\\
&&
\eqa

\bqa
c^{ew}_{+,-,-}(H^-)&=&{\alpha\over4\pi}\{
[-~{1+2c^2_W\over8s^2_Wc^2_W}\log^2\frac{s}{M_W^2}]\nonumber\\
&&+[\log\frac{s}{M_W^2}][-~{5+22 c^2_W\over36s^2_Wc^2_W}+
{m^2_t\over4s^2_WM^2_W}(1+\cot^2\beta)+
{3m^2_b\over4s^2_WM^2_W}(1+\tan^2\beta)]\}\nonumber\\
&&
\eqa

such that
\bqa
F^{Univ}_{-,+,+}&=&
F^{Born}_{-,+,+}[~{\alpha\over4\pi}]\{~[-\log^2\frac{s}{M_W^2}]
[{13+14c^2_W\over36s^2_Wc^2_W}]~\}
\label{funivlsm}
\eqa

\bqa
F^{Univ}_{+,-,-}&=&
F^{Born}_{+,-,-}[~{\alpha\over4\pi}]\{~[-\log^2\frac{s}{M_W^2}]
[{7+20c^2_W\over36s^2_Wc^2_W}]~\}
\label{funivlsm}
\eqa

\underline{b) One loop electroweak angular terms}

\bqa
F^{ang}_{-,+,+}&=&
F^{Born}_{-,+,+}[-~{\alpha\over24\pi c^2_W}][\log\frac{s}{M_W^2}]
\{~[\frac{4}{3}\log{-t\over s}]-{1-10c^2_W\over s^2_W}[\log{-u\over s}]~\}
\eqa
\bqa
F^{ang}_{+,-,-}&=&
F^{Born}_{+,-,-}[-~{\alpha\over12\pi c^2_W}][\log\frac{s}{M_W^2}]
\{~[\log{-u\over s}]-{1\over3}[\log{-t\over s}]~\}
\eqa
\noindent
Again, there are no one loop ew RG terms.
 
\underline{c) One loop SUSY QCD universal terms}

\bqa
F^{Univ~SUSYQCD}_{-,+,+}&=&
F^{Born}_{-,+,+}
[-\frac{\alpha_s}{3\pi}\log\frac{s}{M^2_{\rm SUSY}}]
\eqa

\bqa
F^{Univ~SUSYQCD}_{+,-,-}&=&
F^{Born}_{+,-,-}
[-\frac{\alpha_s}{3\pi}\log\frac{s}{M^2_{\rm SUSY}}]
\eqa

As shown in~\cite{Carena} for the decays $t\to bH^+$, $H^+\to t\overline b$ and 
in~\cite{BGGS,Berger} for this $bg\to tH^-$ process, 
in addition to the SUSY electroweak and QCD
contributions coming from loop diagrams, there are important renormalization terms
in the large $\tan\beta$ limit which affect the Born value Eq.(2.50), namely one should
replace $m_b\tan\beta$ by $m_b\tan\beta/(1+\Delta_b)$. The explicit expression of $\Delta_b$ 
in terms of the various mass and mixing parameters is given in~\cite{Berger}.
This correction becomes important when masses are not degenerate and $\tan\beta$ is 
large, so that we will keep track of this term in the applications that we shall make in 
Section 3.

The (long) list of equations that we have derived in this three 
subsections provide the relevant asymptotic expressions of all the cross 
sections of the four considered processes of single top production.
They represent an original result of our paper, that cannot be compared
with other one loop formulae since these do not exist, to our knowledge.
Note that the Standard Model result can be easily reproduced by our equations
with the simple formal rules that have been already discussed in previous
references and that correspond essentially to the following replacements:
for the gauge part: $2\log(s/M_W^2)-\log^2(s/M_W^2)$
$\to$ $3\log(s/M_W^2)-\log^2(s/M_W^2)$, and for the Yukawa
part: $2m_t^2(1+\cot^2\beta)\to m_t^2$ and 
$2m_b^2(1+\tan^2\beta)\to m_b^2$ in the heavy quark case;
$m_t^2(1+\cot^2\beta)\to m_t^2$ and 
$m_b^2(1+\tan^2\beta)\to m_b^2$ in the Higgs and Goldstone cases.

In our study we have not considered until now the possibility of defining different
observables, e.g. exploiting the fact that the polarization of the final top should be, 
in principle, observable~\cite{CERNYB}. This might lead to the definition of longitudinal
polarization asymmetries already considered in the case of top - antitop production~\cite{ttb}.
As a matter of fact, such observables should not provide
exciting information (apart from possible tests of the V-A structure) for the first two
$t$, $s$-channel processes, where at high energy the final top is supposed to be 
essentially of left handed type. But for the two remaining processes this property
is a priori no longer valid, and the production of a right handed top is not forbidden.
This is made possible by the  nature of the other particles produced in 
association (i.e. either a longitudinal $W$ boson or a scalar charged Higgs),  that breaks 
chirality conservation. We have thus decided to devote the next 
Section 2D to a collection of the
relevant formulae for the related polarization asymmetries.
Again, we remark that, to our knowledge, these formulae do not exist in the literature.
Also, a  preliminary discussion of the expected experimental and theoretical uncertainties
of their measurements, which exists for the four unpolarized production cross
sections~\cite{CERNYB,HiggsWorking} is not yet available. 
Therefore our formulae
should essentially be considered as a theoretical proposal. 
The reason why we decided to show them 
is that we believe that they exhibit some interesting features, that would deserve,
in our opinion, a further investigation performed in a  
more realistic approach.

\subsection{top polarization asymmetries in $t,W^-$ and $t,H^-$
production}

We define in general
\bqa
a_t(\theta)&=&[{d\sigma^{1~loop}(bg\to t_L Y^-)\over d\cos\theta}
-{d\sigma^{1~loop}(bg\to t_R Y^-)\over d\cos\theta}]
/ \nonumber \\
&& [{d\sigma^{1~loop}(bg\to t_L Y^-)\over d\cos\theta}
+{d\sigma^{1~loop}(bg\to t_R Y^-)\over d\cos\theta}]
\eqa
where $Y=W$ or $H$.

In the two considered processes we shall have 
\bq
tW^-:\qquad a_t(\theta) = \frac{|F_{----}|^2+|F_{-+-+}|^2-|F_{+++0}|^2}{|F_{----}|^2+|F_{-+-+}|^2+|F_{+++0}|^2}
\eq
\bq
tH^-:\qquad a_t(\theta) = \frac{|F_{+--}|^2-|F_{-++}|^2}{|F_{+--}|^2+|F_{-++}|^2}
\eq
The asymptotic expansions at one loop of the various helicity amplitudes that enter
equations (2.61-2.62) have been already derived in Sections 2B-C and therefore
we shall not write the one loop expressions of the asymmetries. It might be useful, though,
to observe that $a_t(\theta)$ in both cases is not vanishing at Born level. More 
precisely, we obtain the following expressions
\bq
tW^-:\qquad a_t^{(Born)}(\theta) = \frac{s^2+u^2-\frac{m_t^2}{2M_W^2} t^2}{s^2+u^2+\frac{m_t^2}{2M_W^2} t^2}
\eq
\bq
tH^-:\qquad a_t^{(Born)}(\theta) = \frac{m_b^2\tan^2\beta-m_t^2\cot^2\beta}{m_b^2\tan^2\beta+m_t^2\cot^2\beta}
\eq
Note that both asymmetries are energy independent at the Born level. Their behavior as functions
of $\cos\theta$ (for final $tW^-$) or $\tan\beta$ (for final $tH^-$) 
are shown in Fig.~(\ref{figborn}).

We have now concluded the list of the asymptotic Sudakov expansions of the considered observables of 
single top production. All the formulae have been given at the partonic level.
In the following Section 3, we shall try to derive more realistic predictions for the observable
physical processes.

\section{Applications to observable processes}

\subsection{Unpolarized cross sections}

We shall begin our analysis with the investigation of the electroweak one-loop effects 
in the four considered unpolarized cross sections. With this aim, we shall first 
provide a calculation of the inclusive differential cross section of the actual processes,
defined as usual as ($Y=d$, $\bar b$, $W^-$ or $H^-$):
\bqa
{d\sigma(PP\to t Y+...)\over ds}&=&
{1\over S}~\int^{\cos\theta_{max}}_{\cos\theta_{min}}
d\cos\theta~[~\sum_{ij}~L_{ij}(\tau, \cos\theta)
{d\sigma_{ij\to  t Y}\over d\cos\theta}(s)~]
\eqa
\noindent
where $\tau={s\over S}$, and $(ij)$ represent 
the initial partons of each process. $L_{ij}$ are  the corresponding
luminosities

\bq
L_{ij}(\tau, \cos\theta)=
\int^{\bar y_{max}}_{\bar y_{min}}d\bar y~ 
~[~ i(x) j({\tau\over x})+j(x)i({\tau\over x})~]
\label{Lij}\eq
\noindent
where S is the total pp c.m. energy, and 
$i(x)$ the distributions of the parton $i$ inside the proton
with a momentum fraction,
$x={\sqrt{s\over S}}~e^{\bar y}$, related to the rapidity
$\bar y$ of the $tY$ system~\cite{QCDcoll}.
The parton distribution functions are the 2003 NNLO MRST set available on~\cite{lumi}.
The limits of integrations for $\bar y$ can be written

\bqa
&&\bar y_{max}=\max\{0, \min\{Y-{1\over2}\log\chi,~Y+{1\over2}\log\chi,
~-\log(\sqrt{\tau})\}\}\nonumber\\
&&
\bar y_{min}= - \bar y_{max}
\eqa
\noindent
where the maximal rapidity is $Y=2$, the  
quantity $\chi$ is related to the scattering angle
in the $tY$ c.m.
\bq
\chi={1+\cos\theta\over1-\cos\theta} 
\eq
and 
\bq
\cos\theta_{min,max}=\mp\sqrt{1-{4p^2_{T,min}\over s}}
\eq
expressed in terms of the chosen value for $p_{T,min}$ which gives
the integration limits for  $\cos\theta$ in eq.(\ref{Lij}).\par 

The results of our analysis are shown in Fig.(\ref{figsigmaenergy}) (energy dependence at various $\tan\beta$) and 
(\ref{figsigmatanbeta})
($\tan\beta$ dependence at two special energies $\sqrt{s} = 0.7$ and 1 TeV). The value of $p_{T, min}$
is always chosen to be 50 GeV. 
As one sees clearly from the Figures, there are two main general features that emerge, {\em i.e.}:

a) the one-loop SUSY electroweak effects are systematically large (well beyond the 10 \% level)
in the considered energy range, particularly for the $t$ channel process and for the $tH^-$ process,
where they can reach the 40\% level for $\tan\beta=50$. In this last case, they are as important
as the SUSY QCD renormalization effects considered in~\cite{Berger}.

b) The effects are strongly dependent on $\tan\beta$. They have  a minimum value for $\tan\beta\simeq 10$, then
they increase regularly for either smaller or larger values.

Encouraged by this preliminary information, we have tried to perform a $\chi^2$ fit to 
$\tan\beta$ assuming a reasonably realistic experimental determination of the $\sqrt{s}$
distribution, which also takes into account expected theoretical uncertainties. With this purpose,
we have proceeded in the following way. First, we have separated the first three processes
of single top production ({\em i.e} $td$, $t\bar b$ and $tW^-$) from the fourth $tH^-$ case.
For the first three cases we have followed the pragmatic attitude of assuming, from the general
conclusions given in~\cite{CERNYB}, that the measurements of all the cross sections can 
be performed in the energy range $500-1500$ GeV (with 20 GeV binning) 
with an overall (theoretical and experimental) accuracy of 10 \%.
In fact, an ambitious final goal of 5\% was mentioned in 
these conclusions, so that we might say
to have followed a reasonably conservative attitude.
With this overall uncertainty, we have obtained the results of combined conventional $\chi^2$ fit
that are shown in Fig.(\ref{figchi2}).
In more details, we define  for $Y = d, \overline{b}, W^-$
\bq
\label{chi2a}
{\cal O}_Y(s, \tan\beta) = \frac{d\sigma}{ds}(PP\to t\ Y + \cdots),
\eq
as in Eq.(3.1). Then, for each true value $\tan\beta^*$ of the unknown $\tan\beta$, we compute
\bq
\label{chi2b}
\chi^2(\tan\beta, \tan\beta^*) = \sum_{Y, s}\left(\frac{{\cal O}_Y(s, \tan\beta)-{\cal O}_Y(s, \tan\beta^*)}{\sigma_Y(s, \tan\beta^*)}\right)^2,
\eq
where the sum over $s$ runs over the above range of $\sqrt{s}$ and the assumed experimental error $\sigma_Y$
is fixed by the request of 10 \% accuracy.
Of course, $\chi^2(\tan\beta^*,\tan\beta^*) = 0$. We now vary $\tan\beta$ until we have $\chi^2(\tan\beta, \tan\beta^*) = 1$. 
This determines two values $\tan\beta = \tan\beta^* + \Delta_\pm$ with $\Delta_+>0$ and $\Delta_- < 0$ shown in Fig.~(\ref{figchi2}).

As one sees from Fig.~(\ref{figchi2}), the determination of 
$\tan\beta$ performed in this way might be 
remarkably successful, in particular for the large $\tan\beta\gtrsim 20$ region, which is expected
to be poorly determined at the LHC time~\cite{Abdel}.
Here the error of the fit will be reduced, under our working
assumptions, to the 
small 2-3\% value for $\tan\beta \simeq 50$. Although, we repeat, our
preliminary analysis has been undoubtedly qualitative, we believe that  this result 
could be considered as a first encouraging step.\par
Before presenting our results for the ($t,H^-$) production process, a
few preliminary remarks seem to us to be appropriate. The first one is
that, using the notations of Ref.\cite{Berger}, we have limited our
investigation to the "inclusive" process, i.e. $gb\to tH^-$ (the same
we did in this paper when we considered the inclusive $tW^-$ production). 
In Ref.\cite{Berger}
this process has been determined to NLO in QCD, including the SUSY QCD
corrections, for a general c.m. energy configuration. The results of
the analysis show that the dominant higher order SUSY QCD correction is
not due to virtual loops, but rather to an effect of "coupling
renormalization" type, that can be described by a constant
(c.m. energy independent) parameter defined as $\Delta_b$ (eq.(7) of
that reference), that depends on several MSSM parameters.
For certain choices of the latter (like e.g.
large gluino masses) the effect of $\Delta_b$ on the cross section can
be rather large; more precisely, it can produce modifications in the 
$30-40$ percent range that cannot evidently be neglected and must be
accurately taken into account in any realistic fit to the data.\par
The second remark that we want to make is that the calculation of the
virtual electroweak SUSY corrections to the inclusive ($tH^-$) process
at high energy that we have performed 
is not appearing in Ref.\cite{BGGS,Berger}. 
We repeat again that our analysis is truncated at the next-to leading logarithmic order,
where only the first two (quadratic and linear) terms of a logarithmic
expansion are retained, since for a preliminary indicative
investigation we believe that this approximation should be
sufficient.\par

After these remarks, we are now ready to show the results
of our investigation. Briefly, we assumed an expression for the
inclusive cross section, written in analogy with eq.(3.1), in which
the SUSY corrections contain both the $\Delta_b$ dependent component
of Ref.\cite{BGGS,Berger} and our logarithmic corrections of Sudakov
origin. The correction $\Delta_b$ actually depends on several MSSM parameters.
An inspection of Table I in~\cite{Berger} shows that the largest effects 
are obtained in a region of the parameter space that roughly corresponds to 
large values of $\tan\beta$. In such an extreme case, 
a simple parametrization of $\Delta_b$ is 
\bq
\label{extreme}
\Delta_b=0.004 \tan\beta .
\eq
The coefficient 0.004 is fixed in order to generate a negative correction 
to the cross section of approximately (relative) forty percent for $\tan\beta=50$ 
as shown in Table I of \cite{Berger}. Of course, our choice is rather drastic, but
we used it to show the qualitative features of the competition between
this term and our electroweak components. Adding the electroweak
Sudakov correction, we have repeated the $\chi^2$ analysis performed
in the previous case of the three ($t,d$), ($t,\bar b$), ($t,W^-$)
processes. 

More precisely we have first simulated actual data for $d\sigma/ds$
in the process $bg\to tH^-$ by using our theoretical expression with the full
set of logarithmic Sudakov corrections as well as the term $\Delta_b$.
These {\em fake} experimental data are assumed to have an 
experimental error at the fixed value of 20 \%, a reasonable choice
as supported in~\cite{Berger}. Again, we considered values of $\sqrt{s}$ between 
0.5 and 1.5 TeV with 20 GeV spacing.
As a second step, we made a $\chi^2$ determination of $\tan\beta$. With this aim, we have made
a fit to the simulated experimental data with different theoretical representations
labeled ``$\Delta_b$'', ``e.w.'', and ``e.w.+$\Delta_b$''
according to the set of corrections that are taken into account. The meaning of the three choices
is the following:

\begin{enumerate}
\item $d\sigma^{\rm th}/ds = d\sigma^{\Delta_b}/ds$. We examine what would be the bias and the confidence
bounds in the determination of $\tan\beta$ when the electroweak Sudakov logarithms are not included in the theoretical 
expression.

\item $d\sigma^{\rm th}/ds = d\sigma^{\rm e.w.}/ds$. As before, but retaining in the theoretical expression for the
cross section only the electroweak Sudakov logarithms. This support the statement that the correction $\Delta_b$
is necessary in any realistic investigation in points of the parameter space where the extreme parametrization
(\ref{extreme}) applies.

\item $d\sigma^{\rm th}/ds = d\sigma^{\rm e.w.+\Delta_b}/ds$. In this case, we use as a theoretical representation, the 
expression with which we generated the simulated experimental data. Of course, there is no bias in this case 
and the best estimate of $\tan\beta$ coincides with the {\em true} value used to simulate the experimental data. 
Here, we just compute the limits and compare them to those obtained in the above two cases.
\end{enumerate}

Hence, following the notation of Eq.~(\ref{chi2a}-\ref{chi2b}), we have considered, for this analysis,
\bqa
{\cal O}_H(s, \tan\beta) &=& \frac{d\sigma}{ds}(PP\to t\ H^- + \cdots), \\
\chi^2(\tan\beta, \tan\beta^*)^{\rm kind} &=& \sum_s\left(\frac{{\cal O}_H^{\rm kind}(s, \tan\beta)-{\cal O}^{{\rm e.w.}+\Delta_b}_H(s, \tan\beta^*)}{\sigma_Y(s, \tan\beta^*)}\right)^2,
\eqa
where ``kind'' can be ``$\Delta_b$'', ``e.w.'', and ``e.w.+$\Delta_b$'' as discussed above. Due to
bias in the cases ``kind'' = ``e.w.'' and ``$\Delta_b$'', 
the minimum of $\chi^2$ as a function of $\tan\beta$ at fixed $\tan\beta^*$ is not reached at $\tan\beta = \tan\beta^*$
and also it is not zero, $\chi^2_{\rm min}>0$. As before, we vary $\tan\beta$ until $\Delta\chi^2 = \chi^2 - \chi^2_{\rm min} = 1$.
This determines a confidence interval $\tan\beta^- < \tan\beta < \tan\beta^+$.
In the left part of Fig.(\ref{figchi2th}), we show the curves for $\tan\beta^\pm$ for the various choices
of the theoretical representation of the cross section. In the right part of the same Figure, we plot
instead the percentual value of the ratios  $(\tan\beta^\pm - \tan\beta^*)/\tan\beta^*$ in order to 
emphasize the bias. Of course, in the case ``kind'' = ``e.w. + $\Delta_b$'', there is no bias at all.

From the analysis of Fig.(\ref{figchi2th}), several comments arise.
The various stripes obtained with different theoretical representations of the cross
section have approximately the same width, {\em i.e.} a $\pm 1-2\%$ with respect to the central value.
The bias that is obtained when one of the two kinds of radiative corrections is not included in 
the theoretical representation of the cross section is about $-10\%$ for $\tan\beta=15$
and increases up to about $-20\%$ at $\tan\beta=50$. These values match the corresponding typical
values of the corrections themselves in the cross section.

It seems that we can conclude that the two kinds of corrections, 
in this extreme case, have similar effects and 
neglecting one of them would lead to a large error in the 
estimated central value for $\tan\beta$, even if the 
confidence interval remains small. Of course, 
neglecting at all the two corrections 
would be definitely unacceptable.
If, on the other hand, we accept that the sum of corrections  ``e.w. + $\Delta_b$'' gives
a reliable representation of the data, then we see that it is possible to 
estimate $\tan\beta$ with a rather small error in the range 2-4 \% for all the explored
values of $\tan\beta$ between 15 and 50.
This remarkable feature is clearly a consequence of the fact that the cross section depends
on $\tan\beta$ already at the Born level. 

Of course, one must remember that the precise form of
the correction $\Delta_b$ depends on several MSSM parameters that are in principle
not known. In practice, with no a priori knowledge on the parameters entering
$\Delta_b$ (i.e. sbottom squark masses, gluino mass, $\mu$ parameter, $\dots$), 
the above determination of $\tan\beta$ is expected to give a tight confidence interval of a 
few percent with a systematic bias which is roughly of the same order of the 
partially or totally unknown correction due to $\Delta_b$.

\subsection{Polarization asymmetries}

To conclude our analysis we present now the results of a determination
of the 1-loop SUSY Sudakov effects on the two polarization asymmetries
defined at parton level by eqs.(2.61,2.62). For this preliminary
determination following the treatment for the unpolarized case we
defined the inclusive asymmetries

\bq
A_t(s)=[{d\sigma_{L}\over ds}
-{d\sigma_{R}\over ds}]/[{d\sigma_{L}\over ds}
+{d\sigma_{R}\over ds}]
\eq
with

\bqa
{d\sigma_{L,R}\over ds}&=&
{1\over S}~\int^{\cos\theta_{max}}_{\cos\theta_{min}}
d\cos\theta~[~\sum_{ij}~L_{ij}(\tau, \cos\theta)
{d\sigma_{ij\to  t_{L,R}Y}\over d\cos\theta}(s)~]
\eqa

Figs.(\ref{figatenergy},\ref{figattanbeta}) show the effect of the one loop corrections at variable energy
and variable $\tan\beta$. From these figures one can derive the
following indications:\\

a) in the $(t,W^-)$ case the effects varies, changing sign with 
$\tan\beta$, moving from a positive value of 2 percent for
$\tan\beta=2$ and reaching a negative value of approximately 5 percent
for $\tan\beta=50$.\\

b) in the ($t,H^-$) case, the situation is rather peculiar. The main
feature that emerges is that the value of the asymmetry at Born level
is already $\tan\beta$ dependent and reaches the limiting values $\pm1$
in correspondence to small ($\lesssim 6$) or large $\tan\beta$ values.
In this ranges, one obviously finds that the one loop effect is
practically vanishing. But the interesting fact that one observes is
that, looking at the sign $\pm1$ of the asymmetry, one would be able
to fix the proper range of $\tan\beta$ with extreme reliability. This
property of the inclusive ($t,H^-$) polarization asymmetry has not
been, to our knowledge, stressed in the literature and it seems to us
that it would deserve a more detailed investigation.

\section{Concluding remarks}

The basic assumption of this paper has been a direct discovery of supersymmetry
at LHC (perhaps already at Tevatron), in a moderately light SUSY scenario,
where all sparticle masses $m_i$ satisfy the bound $m_i \le M$, with $M$
of the order of $\sim 350-400$ GeV. Under these conditions, we have
considered, in the theoretical framework of the MSSM, four processes of 
single top production in the kinematical configuration $\sqrt{s}$ (initial
partons center of mass energy) $\sim 1$ TeV, that will be accessible at LHC.
The purpose of our study was that of investigating whether measurements of distributions 
of cross sections or of other observables, performed under realistic theoretical
and experimental precisions, might lead to stringent consistency tests of the 
electroweak sector of the model, at its perturbative one-loop level limit.
Our main tool has been the use of a logarithmic Sudakov expansions for the considered
sector, which appears theoretically justified in the considered scenario at the chosen
c.m. energy. The latter expansion has been computed to next to leading order, 
i.e. retaining the quadratic and linear logarithmic terms and neglecting further
contributions, essentially constant (energy independent) ones.
In such an approach, the only supersymmetric parameter that enters the 
electroweak expansion is $\tan\beta$. Under these conditions, one might hope
that a suitable fit to realistic data can fix with a reasonable accuracy 
the correct $\tan\beta$ value. The outcome might be, depending on the case, 
either a confirmation of a previous alternative determination (e.g. for small
$\tan\beta$ $\lesssim$ 20 values) or possibly a prediction for large $\tan\beta$ values ($\tan\beta > 20$),
where an alternative determination might still be lacking at the LHC running 
time. If the second more appealing and ambitious programme turned out to be the correct one,
the outcome of ``predicting'' the $\tan\beta$ values from precision measurements at LHC
would assume a theoretical relevance comparable, to a certain extent, 
to that belonging to the memorable prediction of the top mass from 
precision measurements at LEP1.

To perform a realistic investigation, i.e. one that might lead
to a measurable prediction for a physical process, a first necessary condition
seemed to us that of examining whether appreciable effects would exist in the basic partonic
processes treated under reasonable qualitative
overall precision assumptions. 
The results of this qualitative investigations show that the overall logarithmic
electroweak effects would be systematically large in the considered $\sqrt{s}$ region
with relative values  between 30-40 \% in all four processes for large $\tan\beta$. 
Starting from this encouraging discovery we have shown that from a combined fit
to the distributions of the cross sections of the first three processes a remarkable ``determination''
of $\tan\beta$ would be possible. For the fourth ($tH^-$) case, we have shown that the situation would
be less simple since a previously determined SUSY QCD constant correction to the Born 
amplitude has to be considered, that would depend on several other parameters of
the model. This would make the fit less simple, although still potentially crucial 
for the large $\tan\beta$ values, for which we believe that this type of searches would be
particularly relevant. In this respect, we have also investigated, although in a more
qualitative an essentially preliminary way, the possible extra information obtainable
from measurements of top polarization asymmetries, finding results that, a priori, 
would seem encouraging, particularly for the previously discussed ($tH^-$)
inclusive production process.

To make the final move to realistic measurable predictions requires now
a few final steps that we list and that we considered, though, beyond
the purposes of this first preliminary investigation. The first step
would be to transform our theoretical one-loop predictions, given (as
it is normally done) in terms of the initial partons c.m. energy $\sqrt{s}$
into predictions given in terms of the final $tY$ pair invariant mass $M_{tY}$.
This requires investigations of the features of, e.g.,  the possible gluon emission 
from the final state that are, in principle,  performable. In fact, at the moment
an analysis of this kind for the case final top antitop production is being carried 
on~\cite{ttb}.
The second step would be that of computing realistic theoretical and experimental
uncertainties for the considered inclusive
distributions. In~\cite{CERNYB,Berger} a preliminary analysis was performed for all
four processes but a rigorous updated systematic and complete 
determination of the theoretical and experimental uncertainties does not exist to 
our knowledge.

A third step would be the extension of our Sudakov expansion to (at least)
the next to next to leading order. This requires the determination of a possible
constant (e.g. energy independent) term which might depend on several parameters of the 
model. An example of this statement has been provided by the consideration of the 
constant $\Delta_b$ term 
in the ($tH^-$) process which depends on the sbottom squarks and gluino masses, on $\mu$ and $A_b$.
Note though, that this $\Delta_b$ term is of \underline{strong} SUSY origin. In our case,
we thus expect a priori that a possible constant \underline{electroweak} SUSY term 
should be depressed by, roughly, a $\sim \alpha/\alpha_s$ factor with respect to $\Delta_b$
which explains why we feel that it should not affect appreciably our preliminary
results, that are due to squared and linear logarithmic enhancements.

As a matter of fact, and as we anticipated at the beginning of this paper, we consider
our long list of equations and our presentation of qualitative figures not really
as final precise predictions, but rather as a proposal for realistic near future
detailed studies. In this sense, we hope that this paper might soon be 
considered as the starting point of future successful and relevant supersymmetry 
studies at LHC.

\newpage

\begin{figure}
\vspace{1.5cm}
\centering
\epsfig{file=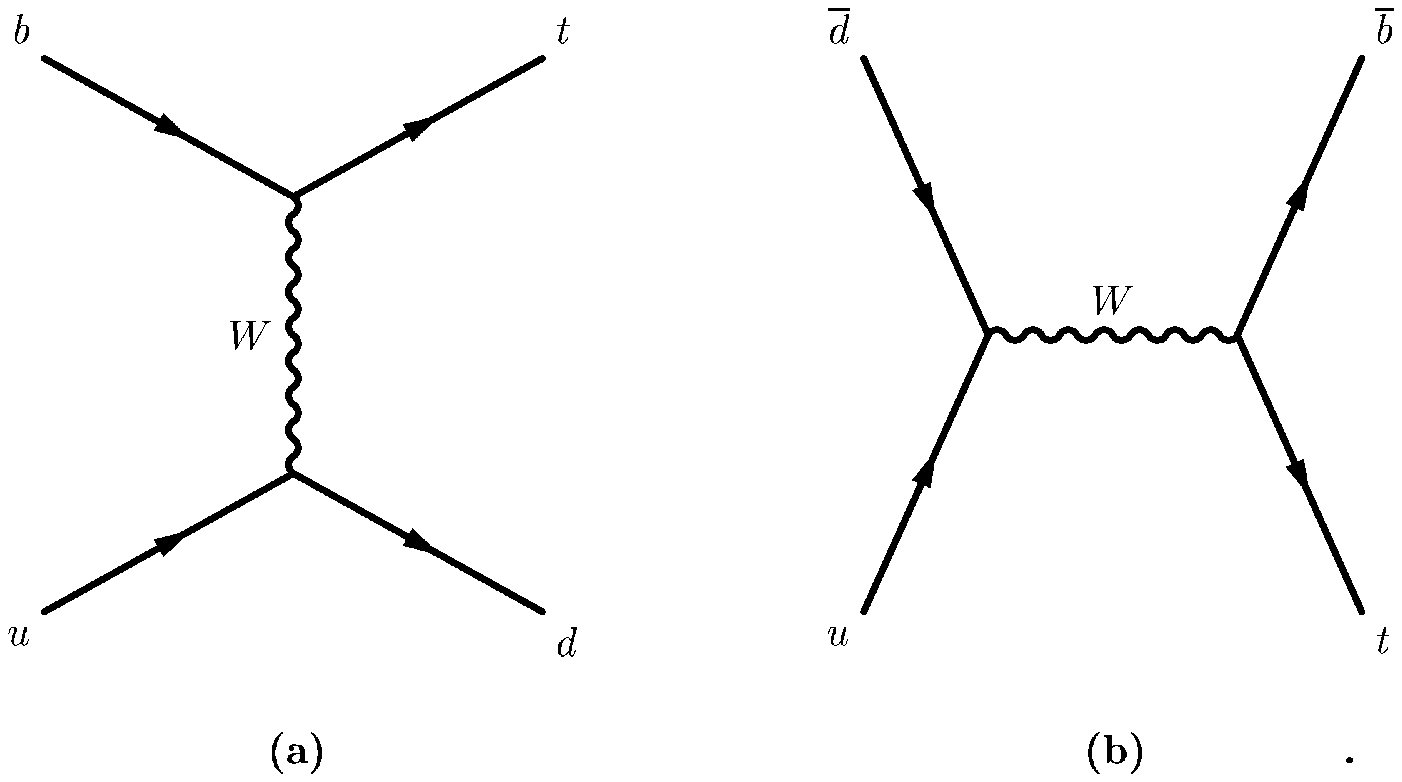,width=12cm}
\vspace{1.5cm}
\caption{Born diagrams for (a) t-channel process $bu\to td$ and
(b) s-channel process $u\bar d\to t\bar b$.}
\label{diagborn1}
\vspace{1.5cm}
\epsfig{file=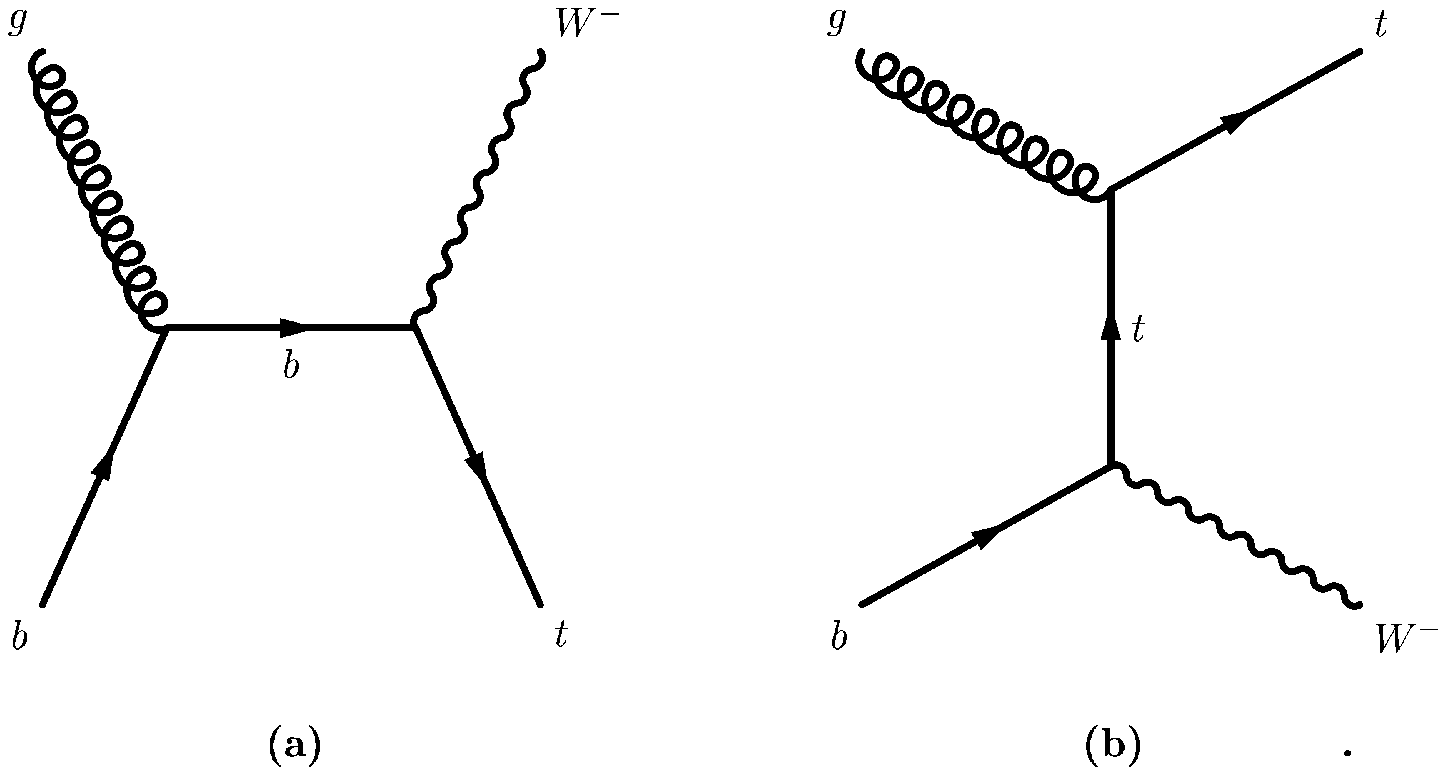,width=12cm}
\vspace{1.5cm}
\caption{Born diagrams for (a) s-channel and
(b) u-channel contributions to the process $gb\to t W^-$.}
\label{diagborn2}
\end{figure}

\newpage

\begin{figure}
\centering
\epsfig{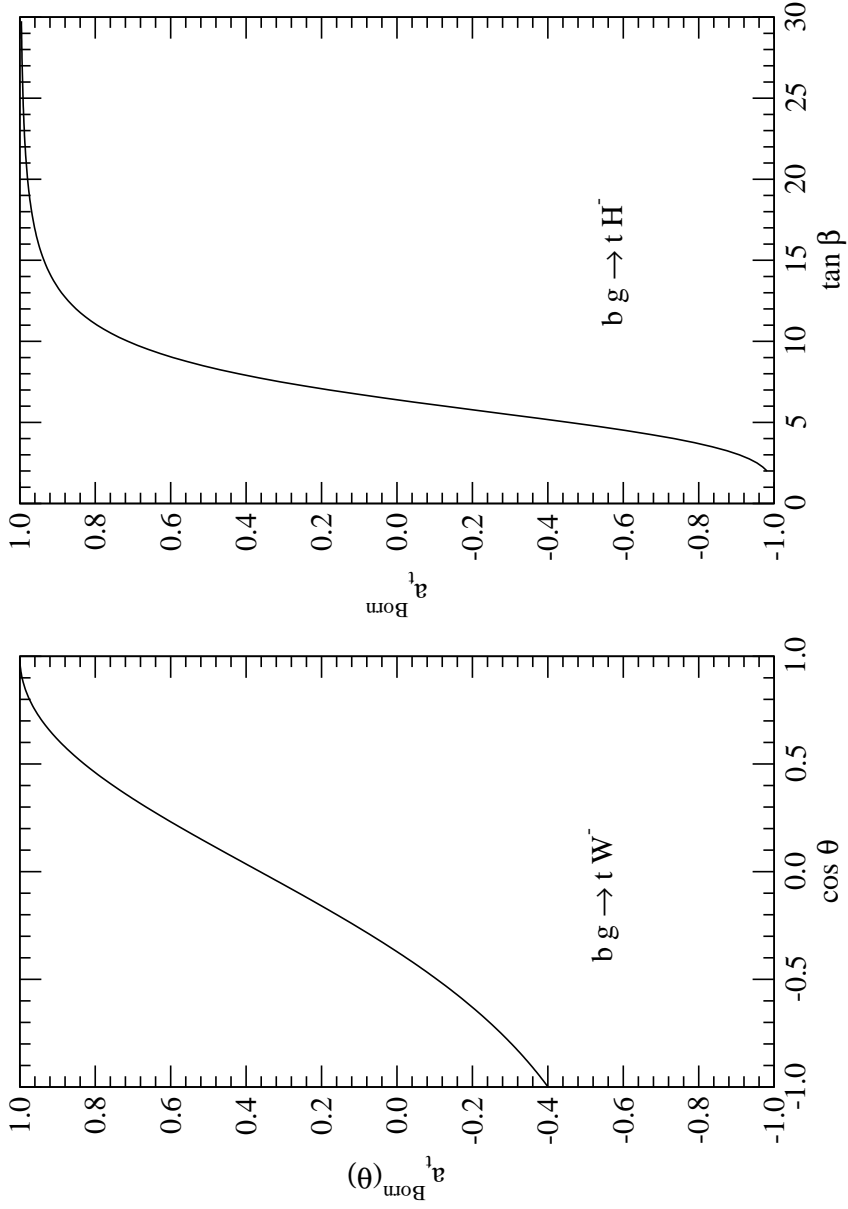}
\vspace{1.5cm}
\caption{Born values of the final top quark polarization asymmetry
for the two processes $bg\to tW^-$ or $tH^-$. Notice that in the first case
the asymmetry is independent on $\tan\beta$, but is scattering angle dependent.
In the second process, these features are reversed.}
\label{figborn}
\end{figure}

\newpage

\begin{figure}
\centering
\epsfig{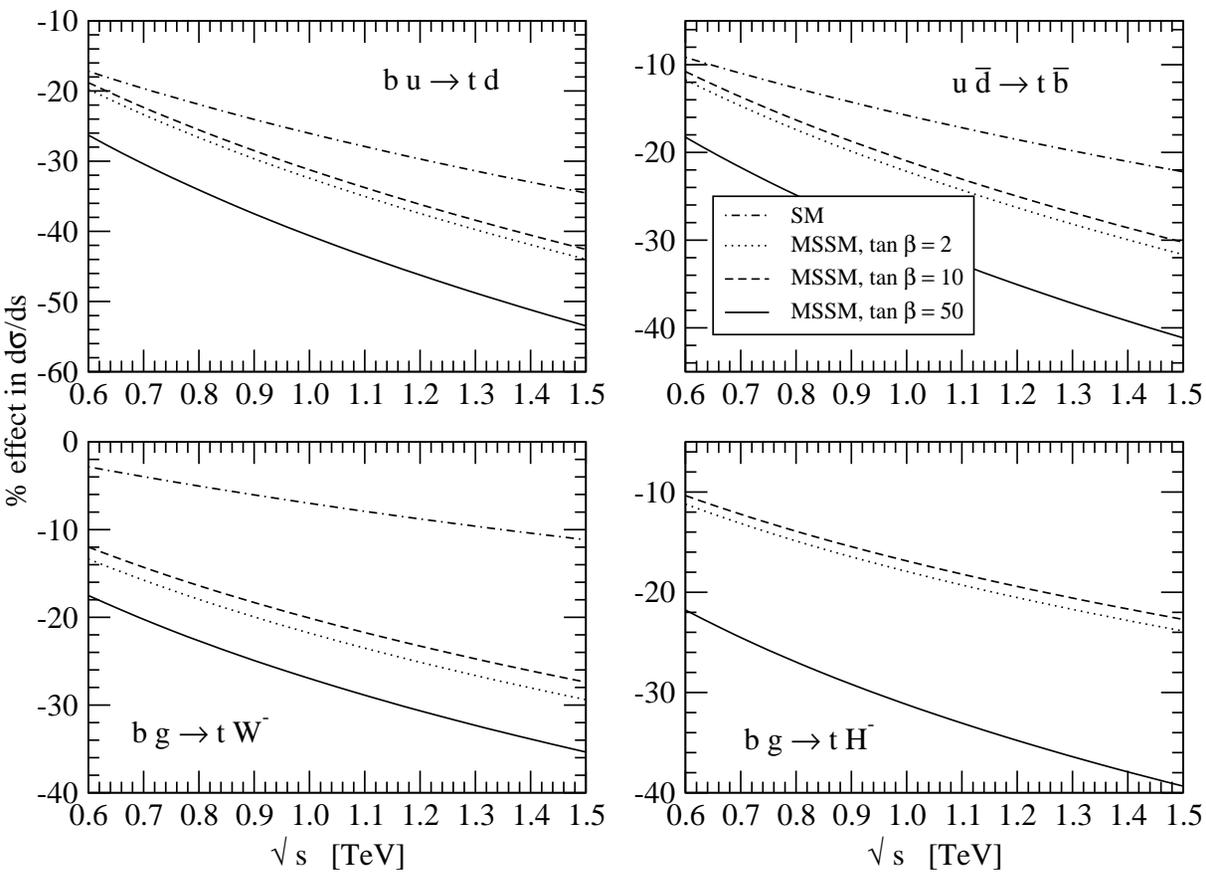}
\vspace{1.5cm}
\caption{Percent correction to the (unpolarized) 
cross section $d\sigma/ds$ due to the full set of Sudakov
logarithmic corrections. We show the curves for the Standard Model
and for the MSSM at three reference values of $\tan\beta$.}
\label{figsigmaenergy}
\end{figure}

\newpage

\begin{figure}
\centering
\epsfig{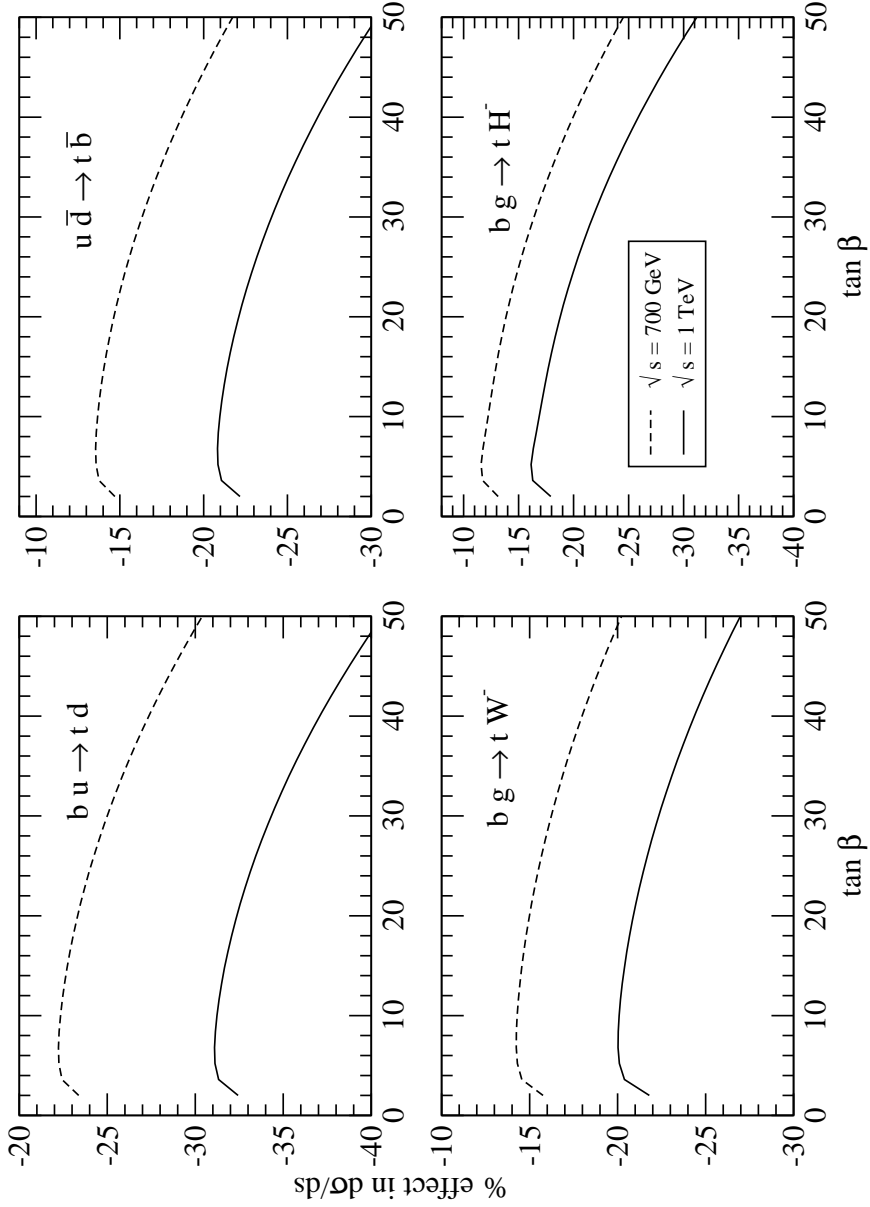}
\vspace{1.5cm}
\caption{Percent correction to the (unpolarized) 
cross section $d\sigma/ds$ due to the full set of Sudakov
logarithmic corrections. We show the curves in the MSSM as functions
of $\tan\beta$ at two reference energy values $\sqrt{s} = 0.7, 1$ TeV.}
\label{figsigmatanbeta}
\end{figure}

\newpage

\begin{figure}
\centering
\epsfig{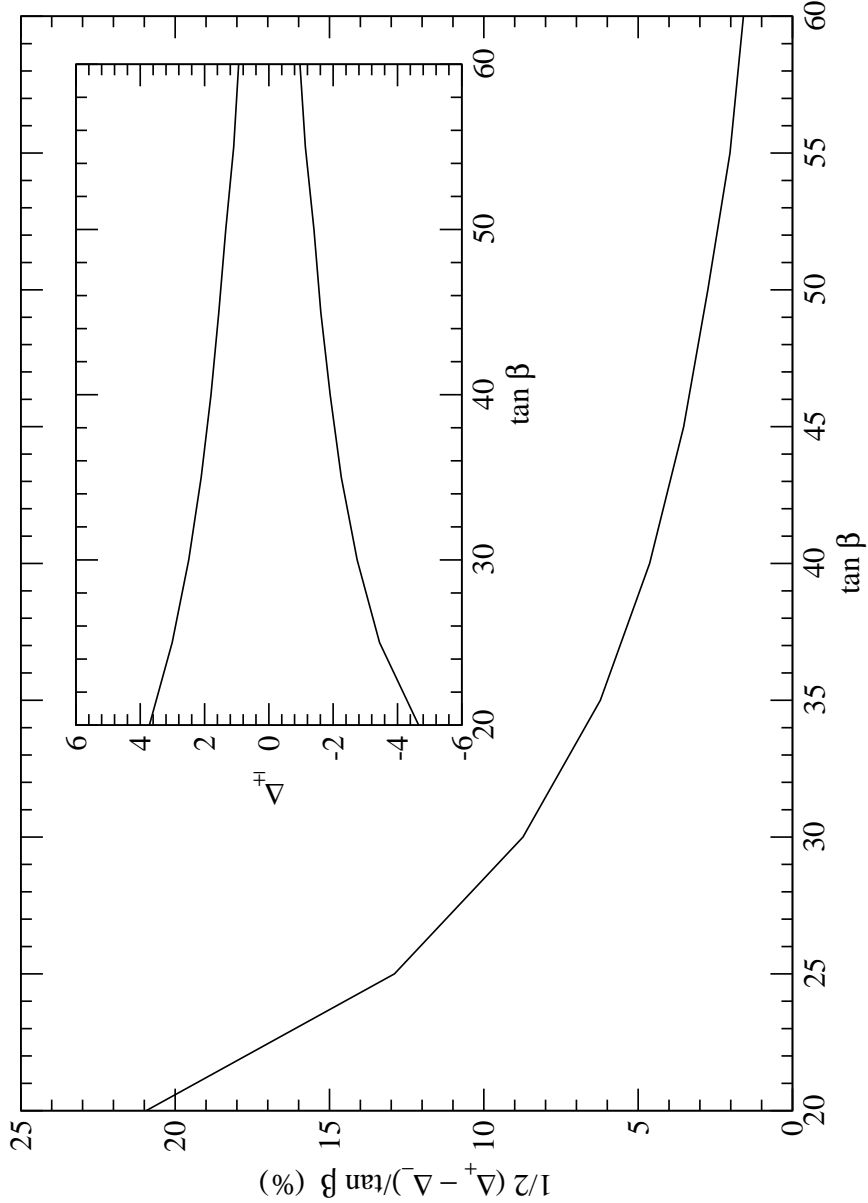}
\vspace{1.5cm}
\caption{Results from the $\chi^2$ analysis of the $\tan\beta$ dependence of the 
cross sections for the processes $bu\to td$, $u\bar d\to t\bar b$ and $bg\to tW^-$.
A 10\% precision is assumed on all the three processes.
For each hypothetical {\em true} value of 
$\tan\beta$, the figures show the $\Delta\chi^2=1$ boundaries $\tan\beta +\Delta_{\pm}$
(shown in the inset plot) and the corresponding relative error 
defined as $\frac{1}{2}(\Delta_+-\Delta-)/\tan\beta$.
The Figure is cut at the upper limit of 25\% since for larger values of the accuracy, the present 
derivation of the confidence bounds deserves a more detailed treatment.
Actually the accuracy is minimum around $\tan\beta\simeq 6$ and is again better than $25\%$ when 
$\tan\beta < 3$.
To avoid confusion we have shown the results of the analysis for $\tan\beta > 20$. }
\label{figchi2}
\end{figure}

\newpage

\begin{figure}
\centering
\epsfig{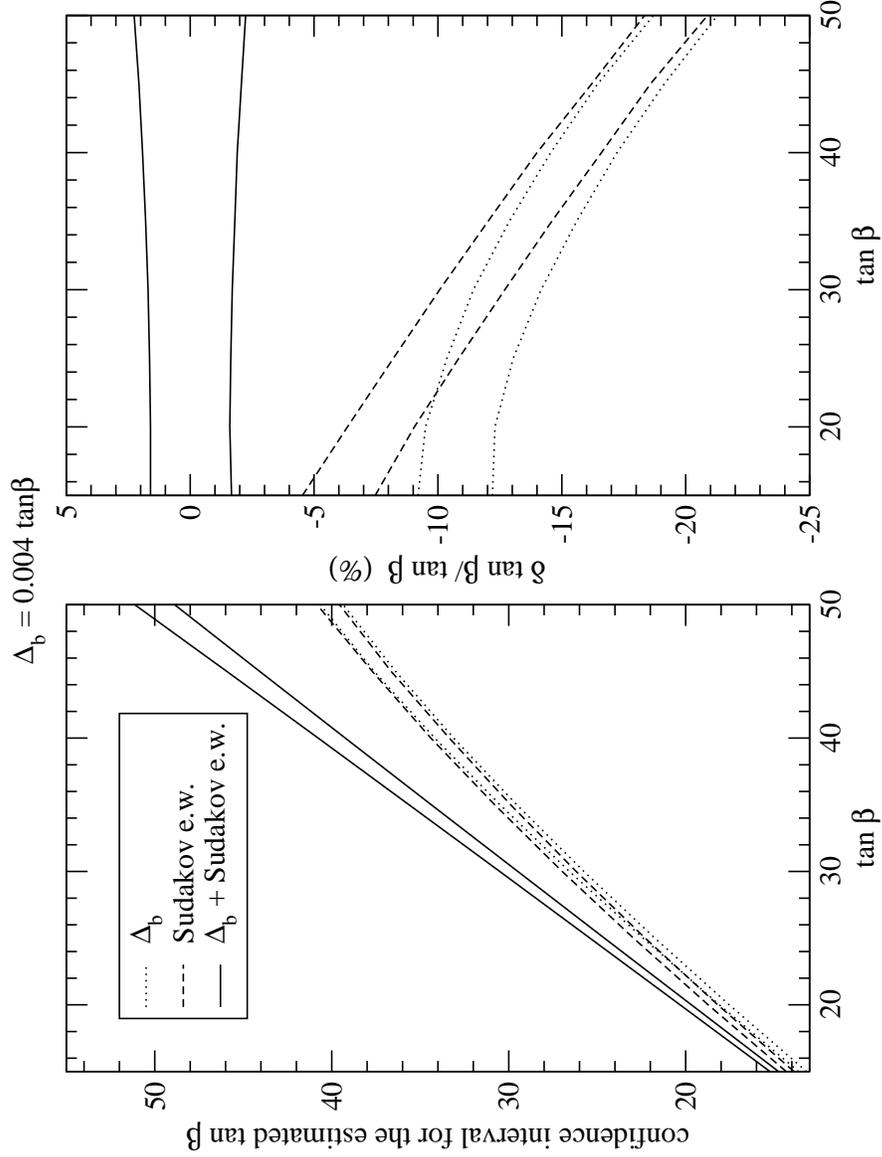}
\vspace{1.5cm}
\caption{Results from the $\chi^2$ analysis of the $\tan\beta$ dependence of the 
cross sections for the process $bg\to tH^-$.
We show the confidence limits $\tan\beta^\pm$ (defined in Sec.~IIIA) on $\tan\beta$ and the 
corresponding relative bounds $(\tan\beta^\pm-\tan\beta)/\tan\beta$
for different choices of the theoretical representation of the cross section
as discussed in the text. In more details, the left figures plots the two curves
corresponding }
\label{figchi2th}
\end{figure}

\newpage

\begin{figure}
\centering
\epsfig{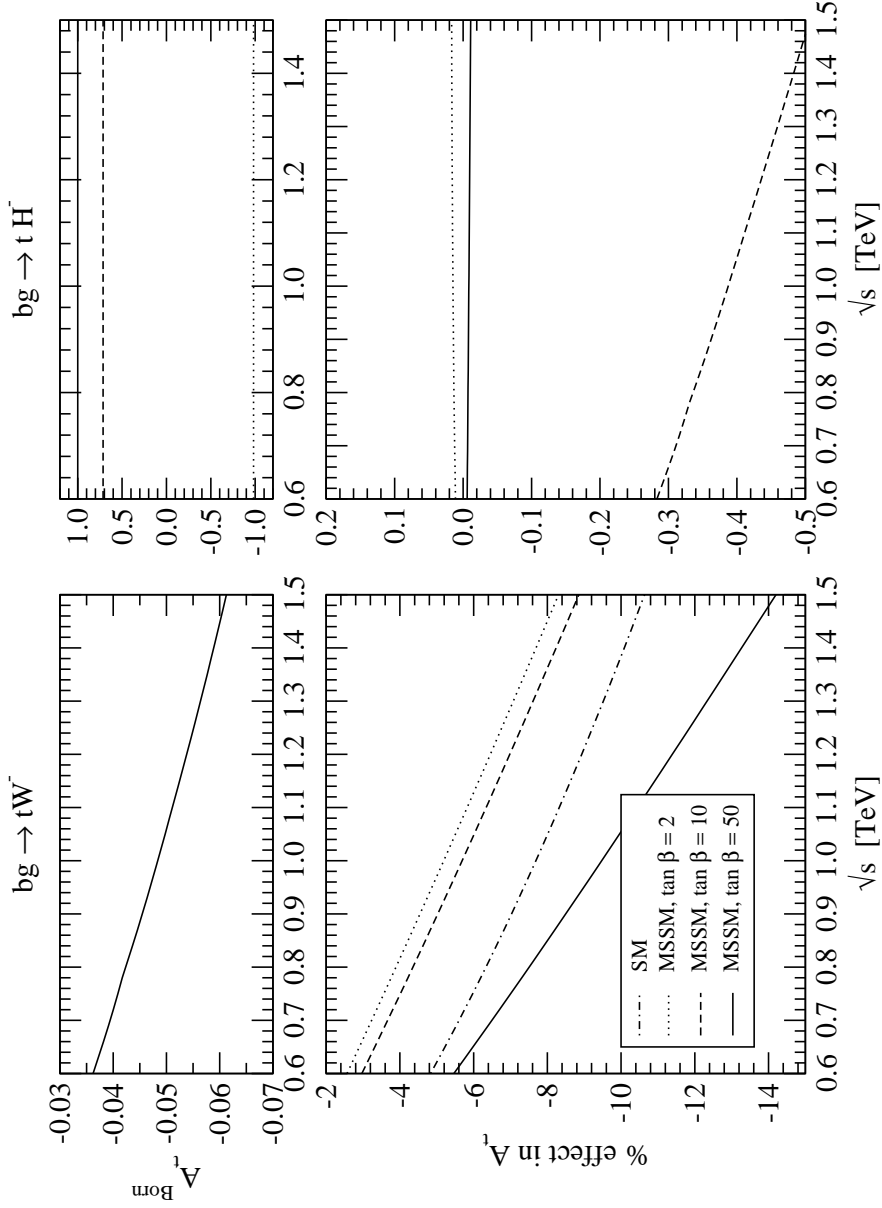}
\vspace{1.5cm}
\caption{Energy dependence of the final top polarization asymmetry.
In the top line, we show for the two processes $bg\to tW^-$, $bg\to tH^-$
the Born values of the asymmetry. In the bottom line, we show the
percentual correction due to the full set of Sudakov
logarithmic corrections. We show the curves for the Standard Model
and for the MSSM at three reference values of $\tan\beta$.}
\label{figatenergy}
\end{figure}

\newpage

\begin{figure}
\centering
\epsfig{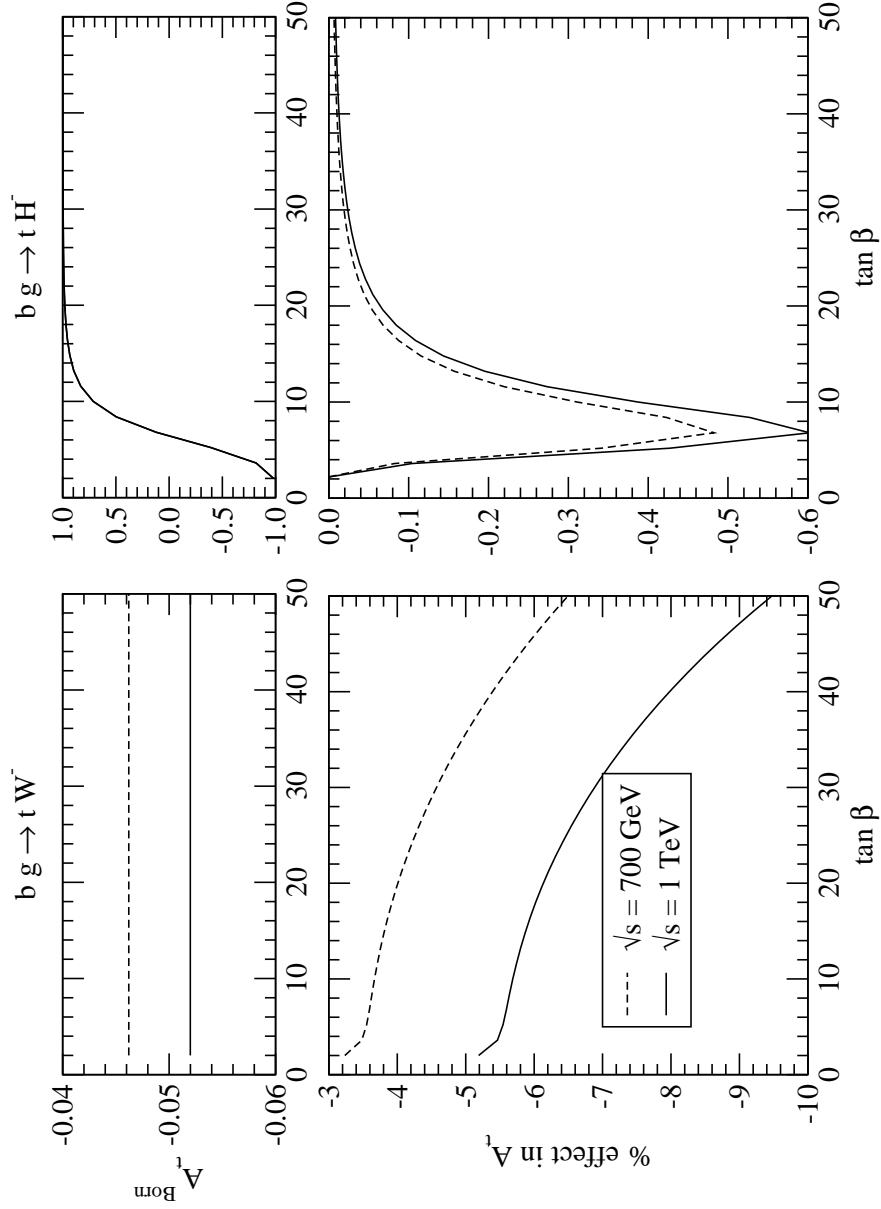}
\vspace{1.5cm}
\caption{
$\tan\beta$ dependence of the final top polarization asymmetry.
In the top line, we show for the two processes $bg\to tW^-$, $bg\to tH^-$
the Born values of the asymmetry. In the bottom line, we show the
percentual correction due to the full set of Sudakov
logarithmic corrections. We show curves for the MSSM 
at two reference energy values $\sqrt{s} = 0.7, 1$ TeV.}
\label{figattanbeta}
\end{figure}

\end{document}